\documentclass[draft=false]{aastex631}

\usepackage{amsmath}
\usepackage[utf8]{inputenc}
\usepackage{xspace}
\usepackage{longtable}

\usepackage{floatrow} 
\usepackage[colorinlistoftodos]{todonotes} 
\usepackage{natbib}

\newcommand {\chandra} {\textsl{Chandra}}
\newcommand{\cxo}{{\it Chandra}\xspace} 
\newcommand {\swift} {\textsl{Swift}}
\newcommand {\nustar} {\textsl{NuSTAR}}
\newcommand {\xmm} {\textsl{XMM-Newton}}
\newcommand {\saao} {\textsl{SAAO}}
\newcommand {\integral} {\textsl{INTEGRAL}}

 \newcommand{\E}[1]{\ensuremath{\times 10^{#1}}}
\def\xsrc{Swift J183920.1$-$045350~}
\def\xsrcns{Swift J183920.1$-$045350}
\def\xabbr{J1839}

\begin{document}


\title{Probing the reach of the Intermediate Polar Cataclysmic Variable Population with Swift\,J183920.1-045350}

\author[0000-0002-1653-6411]{Nicholas M. Gorgone}\email{nicholas.gorgone@gmail.com}
\affiliation{Department of Physics, The George Washington University, Washington, DC 20052, USA}
\affiliation{Astronomy, Physics and Statistics Institute of Sciences (APSIS), The George Washington University, Washington, DC 20052, USA}

\author[0000-0002-6896-1655]{Patrick A. Woudt}
\affiliation{Department of Astronomy, University of Cape Town, Private Bag X3, Rondebosch 7701, South Africa}

\author[0000-0002-7004-9956]{David Buckley} 
\affiliation{South African Astronomical Observatory, P.O. Box 9, 7935 Observatory, South Africa}

\author{Koji Mukai}
\affiliation{CRESST II and X-ray Astrophysics Laboratory, NASA/GSFC, Greenbelt, MD 20771, USA}
\affiliation{Department of Physics, University of Maryland Baltimore County, 1000 Hilltop Circle, Baltimore MD 21250, USA}

\author{Chryssa Kouveliotou}

\author[0000-0002-5274-6790]{Ersin G\"{o}\u{g}\"{u}\c{s}}
\affiliation{Sabanc\i~University, Faculty of Engineering and Natural Sciences, \.Istanbul 34956 Turkey}

\author[0000-0001-8018-5348]{Eric Bellm}
\affiliation{DIRAC Institute, Department of Astronomy, University of Washington, 3910 15th Avenue NE, Seattle, WA 98195, USA}

\author[0000-0002-3873-5497]{Justin D. Linford}
\affiliation{National Radio Astronomy Observatory, P.O. Box O, Socorro, NM 87801, USA}

\author{Alexander J. van der Horst}
\affiliation{Department of Physics, The George Washington University, Washington, DC 20052, USA}
\affiliation{Astronomy, Physics and Statistics Institute of Sciences (APSIS), The George Washington University, Washington, DC 20052, USA}

\author[0000-0003-4433-1365]{Matthew G. Baring}
\affiliation{Department of Physics and Astronomy - MS 108, Rice University, 6100 Main Street, Houston, Texas 77251-1892, USA}

\author[0000-0002-8028-0991]{Dieter Hartmann}
\affiliation{Department of Physics and Astronomy, Clemson University, Kinard Lab of Physics, Clemson, SC 29634-0978, USA}

\author[0000-0002-8456-1424]{Paul Barrett}
\affiliation{Department of Physics, The George Washington University, Washington, DC 20052, USA}

\author{Bradley Cenko}
\affiliation{Astrophysics Science Division, NASA Goddard Space Flight Center, MC 661, Greenbelt, MD 20771, USA}
\affiliation{Joint Space-Science Institute, University of Maryland, College Park, MD 20742, USA}

\author{Melissa Graham}
\affiliation{DIRAC Institute, Department of Astronomy, University of Washington, 3910 15th Avenue NE, Seattle, WA 98195, USA}

\author{Johnathan Granot}
\affiliation{Department of Physics, The George Washington University, Washington, DC 20052, USA}
\affiliation{Department of Natural Sciences, The Open University of Israel, P.O Box 808, Ra'anana 43537, Israel}

\author{Fiona Harrison}
\affiliation{Cahill Center for Astrophysics, California Institute of Technology, 1216 East California Boulevard, Pasadena, CA 91125, USA}

\author{Jamie Kennea}
\affiliation{Department of Astronomy and Astrophysics, The Pennsylvania State University, 525 Davey Lab, University Park, PA 16802, USA}

\author[0000-0002-9700-0036]{Brendan M. O'Connor} 
\affiliation{Department of Physics, The George Washington University, Washington, DC 20052, USA}
\affiliation{Astronomy, Physics and Statistics Institute of Sciences (APSIS), The George Washington University, Washington, DC 20052, USA}

\author{Stephen Potter}
\affiliation{South African Astronomical Observatory, P.O. Box 9, 7935 Observatory, South Africa}

\author{Daniel Stern}
\affiliation{Jet Propulsion Laboratory, California Institute of Technology, 4800 Oak Grove Drive, Mail Stop 169-221, Pasadena, CA 91109, USA}

\author{Ralph Wijers}
\affiliation{University of Amsterdam, Science Park 904, 1098 XH Amsterdam, The Netherlands}

\begin{abstract}
We report on the \swift/XRT Deep Galactic Plane Survey discovery and multi-wavelength follow-up observations of a new intermediate polar Cataclysmic Variable, \xsrcns.
A 449.7~s spin period is found in \xmm{} and \nustar{} data, accompanied by a 459.9~s optical period that is most likely the synodic, or beat period, produced from a 5.6 h orbital period.  The orbital period is seen with moderate significance in independent long-baseline optical photometry observations with ZTF and SAAO. We find that the source X-ray pulsed fraction decreases with increasing energy. The X-ray spectra are consistent with the presence of an Fe emission line complex with both local and interstellar absorption. In the optical spectra, strong H$\alpha{}$, H I, He I and He II emission lines are observed, all common features in magnetic CVs. The source properties are thus typical of known intermediate polars, with the exception of its estimated distance of 2.26$^{+1.93}_{-0.83}$~kpc, which is larger than typical, extending the reach of the CV population in our Galaxy.
\end{abstract}



\section{Introduction}
Cataclysmic Variables (CVs) are binary systems including a white dwarf (WD) accreting material from a companion, usually a main-sequence star, which is undergoing Roche Lobe overflow.
An accretion disk can form
if the magnetic field of the primary (the WD) is sufficiently small, i.e., $B\lesssim{}10^{6}$~G\cite[see e.g.,][]{Warner1992,Warner1995,Coppejans2016}.  Systems with field strengths below this threshold are known as nonmagnetic CVs, whereas the magnetic systems exceeding this field strength are divided into polars and intermediate polars (IPs). The difference between the latter two categories is that in the most magnetic WDs (polars), the Alfv\'{e}n radius (R$_{A}$) extends to the L1 Lagrange point, which stops an accretion disk from forming, while in the lower $B-$field IPs an accretion disk can form between L1 and R$_{A}$. In addition, the magnetic field of polars induces a torque, which approximately synchronizes the WD spins to their orbital periods. In contrast, in IPs the weaker WD fields cannot synchronize spins with orbital periods and the WDs are in fact spun up by accretion torques \citep{Cropper1990,Patterson1994,Wickramasinghe2014}.


In IPs, material at the inner edge of the disk is funnelled along the magnetic field lines towards the poles of the WD, forming a shock in the accretion column as it interacts with the WD atmosphere. Accretion from the inner edge of the disk results in azimuthally extended accretion curtains, akin to the auroral zones on Earth. During accretion, several emission mechanisms are at play; for a comprehensive review see \citet{Mukai2017}. At the shock front, thermal bremsstrahlung is the major contributor to the X-ray emission, while line cooling becomes increasingly important with lower temperatures at the post-shock (due to less-ionized accretion species). Most IP spectra have been fit with reasonable agreement to a bremsstrahlung model plus \textsc{$F_{\rm e}$} lines; however, successful fits to wide-band, high quality X-ray spectra of IPs require more sophisticated models, including multi-temperature plasma models such as \textsc{mkcflow}, complex absorbers, and reflection (see, e.g., \citealt{LopesdeOliveira+2019} and references therein). In a few so-called ``soft" IPs, there is also a soft X-ray component (prevalent in polars) characterized by a blackbody with $T_{bb} \lesssim{}$ 100~eV.

Non-magnetic CVs are most often discovered in the optical, via their variability or their spectral features. Even though the number of known non-magnetic CVs is increasing rapidly, thanks to Sloan Digital Sky Survey (SDSS) and to recent time-domain surveys, such as the All-Sky Automated Survey for Supernovae (ASAS-SN), our census is still only complete within the immediate Solar neighborhood. For example, \cite{Pala+2020} constructed a volume-limited sample of CVs out to 150 pc, which includes SDSS, ASAS-SN, and Gaia discoveries, as well as one object discovered by an amateur astronomer. Magnetic CVs are, in general, the most X-ray luminous sub-class of CVs, and a large majority are discovered as X-ray sources. Polars generally are soft X-ray bright sources, so the current list of polars are dominated by \textsl{ROSAT}-discovered objects (see, e.g., \citealt{Beuermann+2020} and references therein). In contrast, most IPs are hard X-ray bright and soft X-ray faint (the latter being due to strong internal absorption). As a consequence a large majority of currently known IPs have been discovered with the \swift\ BAT and \integral\ hard X-ray surveys \citep{deMartino+2020}. Each new X-ray survey has added to our census of, and our understanding of, magnetic CVs. 

Here we present the discovery with the Niel Gehrels {\it Swift} Observatory Deep Galactic Plane Survey (DGPS; PI C. Kouveliotou) of \xsrc{}(hereafter \xabbr{}); we also present our results of followup observations obtained with a multiwavelength campaign. \xabbr{} is the second source we followed up in-depth within the scope of DGPS \citep{Gorgone+2017}. The DGPS is a {\it Swift} legacy program that covers the inner regions of the Milky Way (Phase I: $30^{\circ{}}>|\ell{}|>10^{\circ{}}$ and $|b|<0.5^{\circ{}}$). In section \ref{sec: observations} we outline the methods used to extract and calibrate multiwavelength data. In section \ref{sec: results} we report on our  sub-arcsecond localization, X-ray and IR spectroscopy, multiwavelength photometry, and timing analysis results. In section \ref{sec: discussion}, we discuss our IP classification of the source. Finally, a summary of our work is presented in section \ref{sec: conclusions}.  




\section{Observations} \label{sec: observations}

\xabbr{} was discovered on 2017, July 13 with the \swift{} DGPS with a 3.8~ks exposure, which revealed a 5.4$\sigma{}$ detection above background ($0.3-10$~keV). However, it was not detected above 3$\sigma$ in the single archival observation of the field with the \swift{} X-Ray Telescope (XRT) in 2013. Since this was a previously uncataloged source, we initiated a series of multi-wavelength observations to identify the optical counterpart and determine the source nature. We observed \xabbr{} with our \chandra{} program Target of Opportunity (ToO) observation for localization as well as with our \xmm{} and \nustar{} ToO programs to obtain X-ray spectra. We also obtained optical spectra with the Dual Imaging Spectrograph (DIS) at the Apache Point Observatory (APO) and with the Robert Stobie Spectrograph (RSS) at the Southern African Large Telescope (SALT), and performed optical photometry with the \swift{} Ultraviolet and Optical Telescope (UVOT) and the \xmm{} Optical monitor (OM). A candidate counterpart was observed in archival data of the Zwicky Transient Facility (ZTF) in the g and r bands; the source flux evolution was tracked through two observing runs covering a total duration of $\sim$700~days. Finally, we took high-speed photometry of the source on 8 nights using the Sutherland High Speed Optical Camera (SHOC) camera on the South African Astronomical Observatory (SAAO) 1-m telescope. In addition we performed a  brief polarimetric observation with the HIgh speed PhotoPOlarimeter (HIPPO). All observations are tabulated chronologically in Table \ref{tbl: observations} and described per wavelength range in detail below.

\subsection{X-ray observations}

\subsubsection{\swift{}/X-Ray Telescope (XRT)}
Count rates were determined using the \textsc{Ximage} routine \texttt{sosta}. We used source regions corresponding to an enclosed-energy fraction of 87\%\footnote{This leads to slightly different region sizes due to the energy dependence of the Point Spread Function.} and local background annuli outside these regions. The count rates were multiplied by a factor (calculated using the \texttt{xrtmkarf} command) to recover the full 100\% of the enclosed-energy fraction. The source was discovered at a count rate of $1.5(3)\times10^{-2}$~counts s$^{-1}$ in photon counting (PC) mode. We also used data from the only archival observation taken in March 2013 with \swift{}/XRT in PC mode. 

\subsubsection{\chandra{}}
We used one of our approved Targets of Opportunity to observe \xabbr{} with the \chandra{} ACIS-I \citep{Garmire+2003} for 2.5~ks on  2018, November 9 (Obs. 9 in Tbl. \ref{tbl: observations}). To prepare the data for analysis, we used routines \texttt{fluximage} and \texttt{dmcopy} from the \texttt{ciao v4.9} package. We determined the location of \xabbr{} with sub-arcsecond accuracy (see section \ref{sec: localization}).

\subsubsection{\xmm{}}
We used one of our approved Targets of Opportunity to observe \xabbr{} for 26~ks on 2018, October 18 with the \xmm{} \citep{Jansen+2001} EPIC cameras \citep[PN; MOS1/2 -][]{Struder+2001, Turner+2001} in full frame imaging mode. To extract the data, we used the Science Analysis Software \texttt{SAS v.1.2} \citep{Gabriel+2004}. We produced the final event files with \texttt{epchain} and \texttt{emchain} commands. We created circular source regions that contained 80\% of the enclosed energy, with radii of 35\arcsec{} and 25\arcsec{} for the PN and MOS cameras, respectively\footnote{See \S{} 3.2.1.1 of the XMM Users Handbook \href{https://heasarc.nasa.gov/docs/xmm/uhb/onaxisxraypsf.html}{https://heasarc.nasa.gov/docs/xmm/uhb/onaxisxraypsf.html}}. We extracted background regions of the same size from nearby, source-free regions. We filtered out times of high background in the 10--12~keV band and only included event patterns 0--4 for PN and 0--12 for both MOS lightcurves. Finally we corrected the event times to the Solar System barycenter using \texttt{barycen} with the source RA and DEC as found in our \chandra{} observation. We then filtered the master events file by energy to obtain data in three bands, namely $0.3-3.0$~keV, $3.0-10.0$~keV and $0.3-10.0$~keV, for our timing and spectral analyses.

\subsubsection{\nustar{}}
We observed \xabbr{} with \nustar{} \citep{Harrison+2013} and used \texttt{heasoft v.6.23} command \texttt{nuproducts} to produce level 3 data. We used the same circular source region (with a radius of 120\arcsec{}) centered on the \cxo{} location, in both focal plane modules. For the background, we used a circular region of the same radius from a source-free field. We barycenter-corrected the photon arrival times to the Solar System based on the \cxo{} position (see \autoref{sec: localization}). We truncated the  \nustar{} data included in our fits at 20~keV, where the background flux started to dominate the source flux. 

\subsection{UV observations}

Two of the X-ray satellites we used are equipped with ultraviolet (UV) monitors and provided contemporaneous UV and X-ray observations.


\subsubsection{\xmm{}/Optical Monitor}
We observed \xabbr{} with the UVM2 ($\lambda{}_{eff}=231$~nm), UVW1 ($\lambda{}_{eff}=291$~nm), and U ($\lambda{}_{eff}=344$~nm) filters of the \xmm{}/OM \citep[][]{Mason+2001}. We used the \texttt{SAS} command \texttt{omichain} to process the data, which aggregated a final list of uniquely detected sources for each filter and produced photometric measurements for each source. 

\subsubsection{\swift{}/Ultraviolet-Optical Telescope}
The \swift{}/UVOT \citep[][]{Roming+2005} is identical to the  \xmm{}/OM\footnote{\swift{}/UVOT modules are the flight spares of the \xmm{}/OM., see \href{https://swift.gsfc.nasa.gov/about\_swift/uvot\_desc.html}{https://swift.gsfc.nasa.gov/about\_swift/uvot\_desc.html}}, with slightly different filter throughput. We observed \xabbr{} with the UVM2 ($\lambda{}_{eff}=225$~nm), UVW1 ($\lambda{}_{eff}=268$~nm), and U ($\lambda{}_{eff}=352$~nm) filters. To process these data, we used the level 3 stacked sky images produced by the HEASARC and used the \textsc{Heasoft} command \texttt{uvotsource} with a 3\arcsec{} radius region, centered on the \chandra{} location. A nearby source-free region of radius 16\arcsec{} was extracted as a background reference.

\subsection{Optical observations}

\subsubsection{Zwicky Transient Facility photometry}
We retrieved publicly-available photometry of \xabbr{} from the Zwicky Transient Facility \citep[ZTF;][]{Bellm+2019,Graham:19:ZTFScience} Data Release 4\footnote{\url{https://www.ztf.caltech.edu/page/dr4}}.  
ZTF observed \xabbr{}
sporadically between 2018, March 28 to 2020, June 28, a total of 135 times with the ZTF g band filter and 241 times with the ZTF r band filter. PSF photometry was automatically extracted using the pipeline described in \citet{Masci+2019}. We found one source in ZTF data within the \chandra{} error circle.


\subsubsection{Southern African Large Telescope (SALT) spectroscopy}
Spectroscopy of \xabbr{} was undertaken with the Southern African Large Telescope \citep[SALT;][]{Buckley2006SPIE.6267E..0ZB} during one night in 2019 and six in 2020 (see Table \ref{tbl: observations}). The Robert Stobie Spectrograph \citep[RSS;][]{Burgh2003SPIE.4841.1463B} was used, initially with the PG900 VPH grating, covering the region 4060--7120~\AA{} at a mean resolution of 4.7~\AA{} with a 1.5\arcsec{} slit width. All exposures were 1800 s. The last three observations utilized the PG1800 VPH grating, covering 5800--7100~\AA{} at a resolution of 2.4~\AA{}, also with a 1.5 \arcsec{}  slit. Six repeat 500 s exposures were taken. For all observations, wavelength calibration lamp exposures were taken immediately following the observations on each night.

The spectra were reduced using the PySALT package
\citep{Crawford2010SPIE.7737E..25C}\footnote{\url{https://astronomers.salt.ac.za/software/pysalt-documentation/}}, which does bias, gain and amplifier cross-talk corrections, mosaics the three CCDs and applies cosmetic corrections. Object extraction, wavelength calibration and background subtraction were all done using standard IRAF\footnote{\url{https://iraf.noao.edu/}} routines, as was the relative flux calibration. 


\subsubsection{South African Astronomical Observatory (SAAO) photometry and photopolarimetry}
Time series photometry of \xabbr{} was undertaken on 8 nights (see Table \ref{tbl: observations})  using the SAAO 1~m telescope\footnote{See \href{https://www.saao.ac.za/astronomers/telescopes-1-0m/}{https://www.saao.ac.za/astronomers/telescopes-1-0m/} for telescope details.} with the Sutherland High speed Optical Camera (SHOC) CCD camera. SHOC uses an Andor iXon888 frame transfer EM-CCD frame camera, with 1024 $\times$ 1024 pixels \citep{Coppejans2013PASP..125..976C}, in conventional (non electron-multiplying; EM) mode. All observations were done without a filter (i.e., `white light'); an exposure time of 30 sec was used. 

Reduction of the CCD images included subtraction of median bias and flat-field correction using median-combined frames from exposures of the twilight sky. Aperture-corrected photometry was used to extract the light curves of all stars in the calibrated science images and differential photometry was performed using several local reference stars. 

Time resolved filterless all-Stokes polarimetry of \xabbr{} was obtained on 2020, July 15, over a period of $\sim$1400 s with the HIPPO photopolarimeter \citep{Potter2010}.


\subsection{Radio observations}

\subsection{Karl G. Jansky Very Large Array (VLA)} 
We were allocated 6 hours of Director's Discretionary Time (DDT) to observe \xabbr{} with the Karl G. Jansky Very Large Array (VLA) in the X-band (8--12~GHz). X-band was chosen to maximize sensitivity, while decreasing Radio Frequency Interference (RFI). Despite the band selection, both observations were significantly affected by RFI, and we removed bands above 10.8~GHz to mitigate spurious signals. All VLA observations used the 3-bit continuum mode with 4 GHz of continuous bandwidth.  In our first observation (Obs. 15 in Table \ref{tbl: observations}), \xabbr{} was observed with 27 antennas. In Obs. 16 and 17, 19 and 25 antennae were used, respectively. Throughout all of our observations, we used J1832-1035 as the complex gain calibrator. We used 3C48 (J0137+3309) as the flux density and bandpass calibrator. We note that 3C48 was known to undergo flaring activity at the time of the observation, causing the absolute flux density estimates to be uncertain to $\sim10\%$.

\begin{table}[]
\begin{tabular}{lclccc}
\hline
Obs. & ID & Telescope & Instrument/Mode & Start Time [UT] & Duration \\
     &    &           &                 & [dd Mmm yyyy hh:mm] & [ks] \\ \hline\hline
1.  & 00044416001 & \swift{}  & XRT/PC + UVOT   & 21 Mar. 2013 06:59   &  0.5 \\
2.  & 00087393001 & \swift{}  & XRT/PC + UVOT   & 13 Jul. 2017 06:02   &  3.8 \\
3.  & 00087393002 & \swift{}  & XRT/PC + UVOT   & 14 Nov. 2017 02:42   &  1.1 \\
4.  & 00010900001 & \swift{}  & XRT/PC + UVOT   & 01 Oct. 2018 17:09   &  1.0 \\
5.  & 0821860201  &\textsl{XMM} & PN + MOS + OM & 18 Oct. 2018 11:27   & 26.0 \\
6.  & 30360002002 & \nustar{} & FPMA/B      & 02 Nov. 2018 08:04   & 40.8 \\
7.  & 00088814001 & \swift{}  & XRT/PC + UVOT   & 02 Nov. 2018 12:30   &  1.4 \\
8.  & 00088814002 & \swift{}  & XRT/WT + UVOT   & 08 Nov. 2018 13:34   &  0.6 \\
9.  & 20335       & \cxo{}    & ACIS-I          & 09 Nov. 2018 08:49   &  2.5 \\
10. & 00087393003 & \swift{}  & XRT/PC + UVOT   & 07 Jul. 2019 21:29   &  0.2 \\
11. &    -        & APO       &    DIS/red      & 13 Jul. 2019 05:54        &  3.6\\
12. &    -        & SALT      &    RSS PG900         & 22 Jul. 2019 19:23   &  1.8\\
13. &    -        & SAAO 1-m      &     SHOC, Clear  &  9 Oct. 2019         &  7.3\\
14. &    -        & SAAO 1-m      &     SHOC, Clear  & 13 Oct. 2019         &  7.2\\
15. & 19B-340     & VLA       &X-band, D config.& 31 Dec. 2019 19:16   &  7.3\\
16. & 19B-340     & VLA       &X-band, D config.& 03 Jan. 2020 19:14   &  7.3\\
17. & 19B-340     & VLA       &X-band, D config.& 04 Jan. 2020 19:28   &  7.2\\
18. &    -        & ZTF       &r-band \& g-band & continuous           & - \\
19. &    -        & SAAO 1-m      &     SHOC, Clear           & 22 May 2020 23:16          &  15.3  \\
20. &    -        & SAAO 1-m      &     SHOC, Clear           & 15 Jul. 2020 20:14        &  6.1   \\
21. &    -        & SAAO 1.9-m      &     HIPPO, Clear           & 16 Jul. 2020 00:14        &  1.4   \\  
22. &    -        & SAAO 1-m      &     SHOC, Clear           & 16 Jul. 2020 18.31        &  25.3   \\
23. &    -        & SAAO 1-m      &     SHOC, Clear           & 21 Jul. 2020 19:03        &  14.9  \\
24. &    -        & SALT      &    RSS PG900          & 21 Jul. 2020 19:45         & 1.8  \\
25. &    -        & SALT      &    RSS PG900         & 23 Jul. 2020 19:17         & 1.8 \\
26. &    -        & SALT      &    RSS PG900         & 24 Jul. 2020 18:58        & 1.8 \\
27. &    -        & SALT      &    RSS PG1800         & 13 Aug. 2020 21:18        & 0.5 $\times$ 6 \\
28. &    -        & SAAO 1-m      &     SHOC, Clear           & 19 Aug. 2020              & 12.0      \\
29. &    -        & SAAO 1-m      &     SHOC, Clear           & 22 Aug. 2020              & 14.6      \\
30. &    -        & SALT      &    RSS PG1800         & 22 Aug. 2020 20:33        & 0.5 $\times$ 6 \\
31. &    -        & SALT      &    RSS PG1800         & 19 Sep. 2020 18:54        & 0.5 $\times$ 6 \\

\hline\hline
\end{tabular}
\caption{Multi-wavelength Observations of  \xsrc{} are enumerated in chronological order.}

\tablecomments{ $\star{}$~No HR calculation possible (source upper limit), $\dagger{}$~Source outside field of view or in damaged pixel area, - Outside of the $0.3-10.0$~keV soft X-ray window.}
\label{tbl: observations}
\vspace{0.5cm}
\end{table}


\section{Results} \label{sec: results}
Here we describe the source localization, and the temporal and spectral analyses results. Contrary to the previous section, which was organized by wavelength, here subsections are organized according to the type of analysis performed. A comprehensive summary of source characteristics can be found in the final table at the discussion part of this paper (Table \ref{tbl: summary parameters}).

\subsection{Localization, Distance, and Proper Motion} \label{sec: localization}
\begin{figure}[!h]
    \centering
    \includegraphics[scale=0.5]{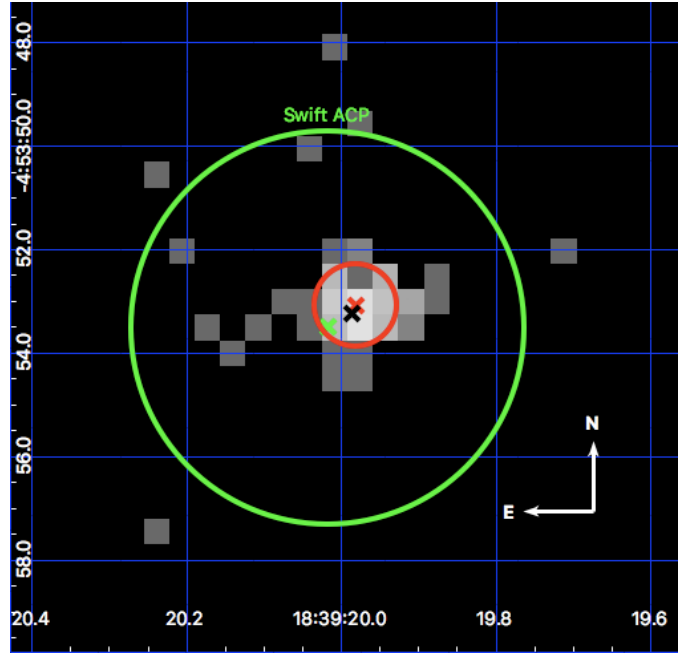}
    \caption{Chandra image (Obs.\,9 from Tbl.\,\ref{tbl: observations}) with the \texttt{wavdetect} position and pointing uncertainty ($\sim{}0.8$\arcsec{}) in red, the XRT/ACP position with a 90\% uncertainty circle from Obs.\,2 (green), and the Gaia location (black x, uncertainty of $\sim{}0.2~mas$). The image is binned at the native plate scale of \chandra{}, 1~pixel$\sim{}$0.5\arcsec{}.}
    \label{fig: cxoposition}
\end{figure}

We determined the \chandra{} location of \xabbr{} using \texttt{wavdetect} with \texttt{ciao v4.9} (Obs.\,9 in Tbl.\,\ref{tbl: observations}). We found one source at R.A. 18$^{h}$39$^{m}$19$^{s}$.98, decl. -04$^{\circ{}}$ 53\arcmin{} 53.1\arcsec{}  (J2000) with a positional uncertainty of 0.8\arcsec{} (90\% confidence, systematic error) within the astrometrically-corrected XRT-UVOT position (ACP). The latter was calculated using the method described in \cite{Evans+2014}. Fig.\,\ref{fig: cxoposition} shows the \chandra{} field with the XRT/ACP error circle (green) superposed on the \chandra{} source (red). The XRT/ACP location center is offset from the \cxo{} location by 0.683\arcsec{}.


Further, we searched the Gaia DR2 data \cite{Gaia+2018} within the \cxo{} uncertainty region. We identified one source, Gaia DR2 4256603449854150016 ($\text{G}_{Gaia}$=18.5), which is offset from the \chandra{} location centroid by 0.193\arcsec{} (black cross, Fig. \ref{fig: cxoposition}). Taking into account the significantly larger \swift{}/XRT{} point spread function, we conclude that the \chandra{} and Gaia sources are indeed the counterparts of  \xabbr{}. 

Finally, we adopt the distance and proper motion of the Gaia source for \xabbr{} for the remaining analysis. The former is estimated to be 2.26$^{+1.93}_{-0.83}$~kpc \cite[68\% confidence; ][]{Bailer-Jones+2018}. The proper motion is $\mu{}_{\alpha{}}$, $\mu{}_{\delta{}}=-1.1 \pm{} 0.4$,$ -2.6 \pm{} 0.4$~mas yr$^{-1}$, where the uncertainties are expressed as standard errors by the \cite{GaiaCollab2018}. In 
Section \S{}\ref{sec:photometry} we discuss the optical and IR photometry results within the source distance context.

\subsection{Spectroscopy} \label{sec:Spectroscopy}

We observed \xabbr{} in three different wavebands: in the X-rays ($0.5-20$~keV), and the optical between  $4000-7000$~\AA{} and $5500-9000$\AA{}. Below we describe our spectral analyses.


\subsubsection{X-ray Spectrum} \label{ssec:Xray Spectrum}

\begin{figure}[!h]
    \centering
    \includegraphics[width=0.98\textwidth, draft=false]{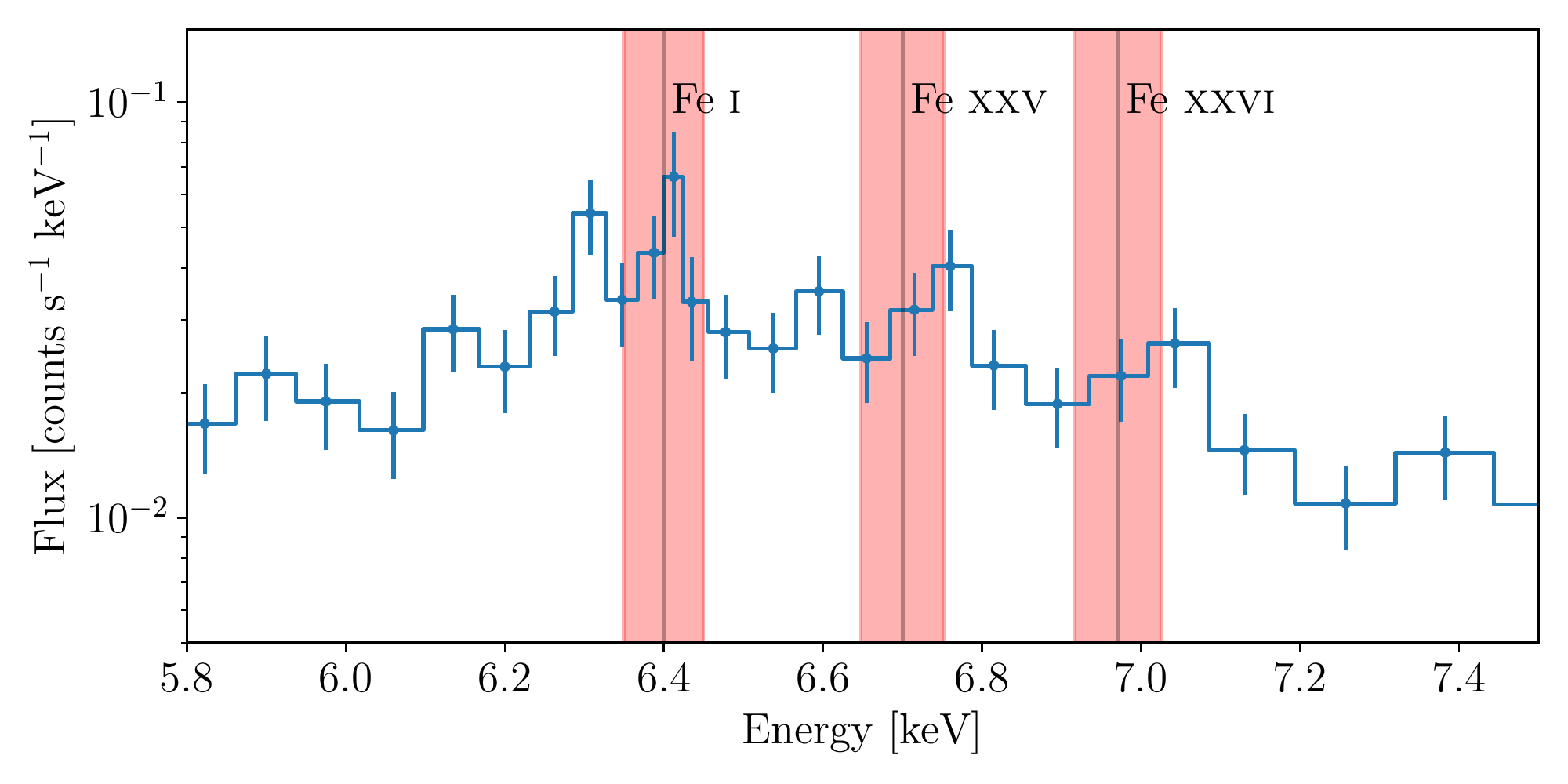}
    \caption{Spectral lines as observed with \xmm{} EPIC PN are shown with the rest energies of the Fe species. Red shaded regions show possible energy shifts in the lines if the source has a velocity along the line of sight of $|v_{r}|\leq{}2.34\times{}10^{3}$~km s$^{-1}$.}
    \label{fig:pn Fe lines}
\end{figure}

\begin{figure}[!h]
    \centering
    \includegraphics[width=0.85\textwidth, draft=false]{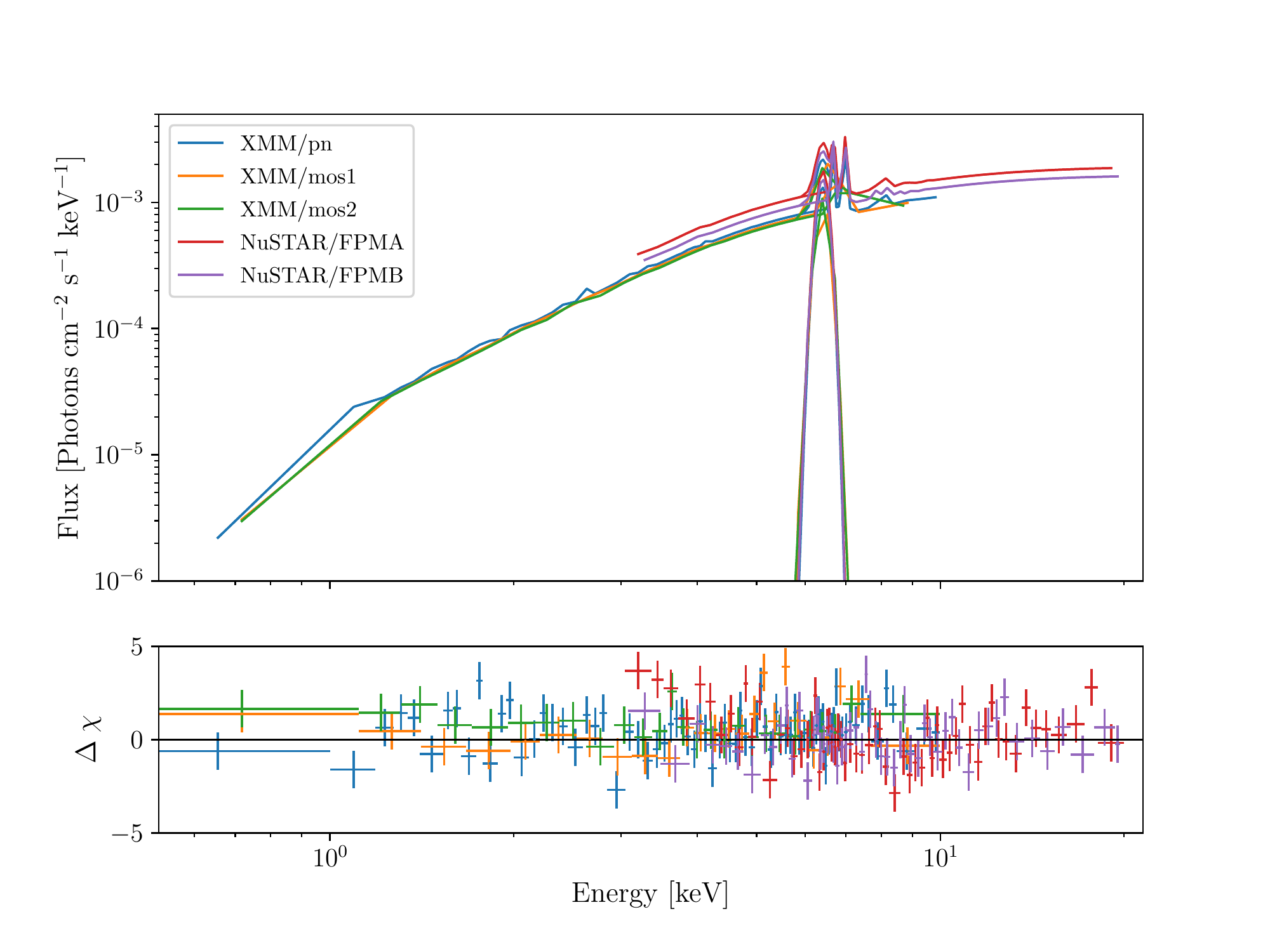}
    \includegraphics[width=0.85\textwidth, draft=false]{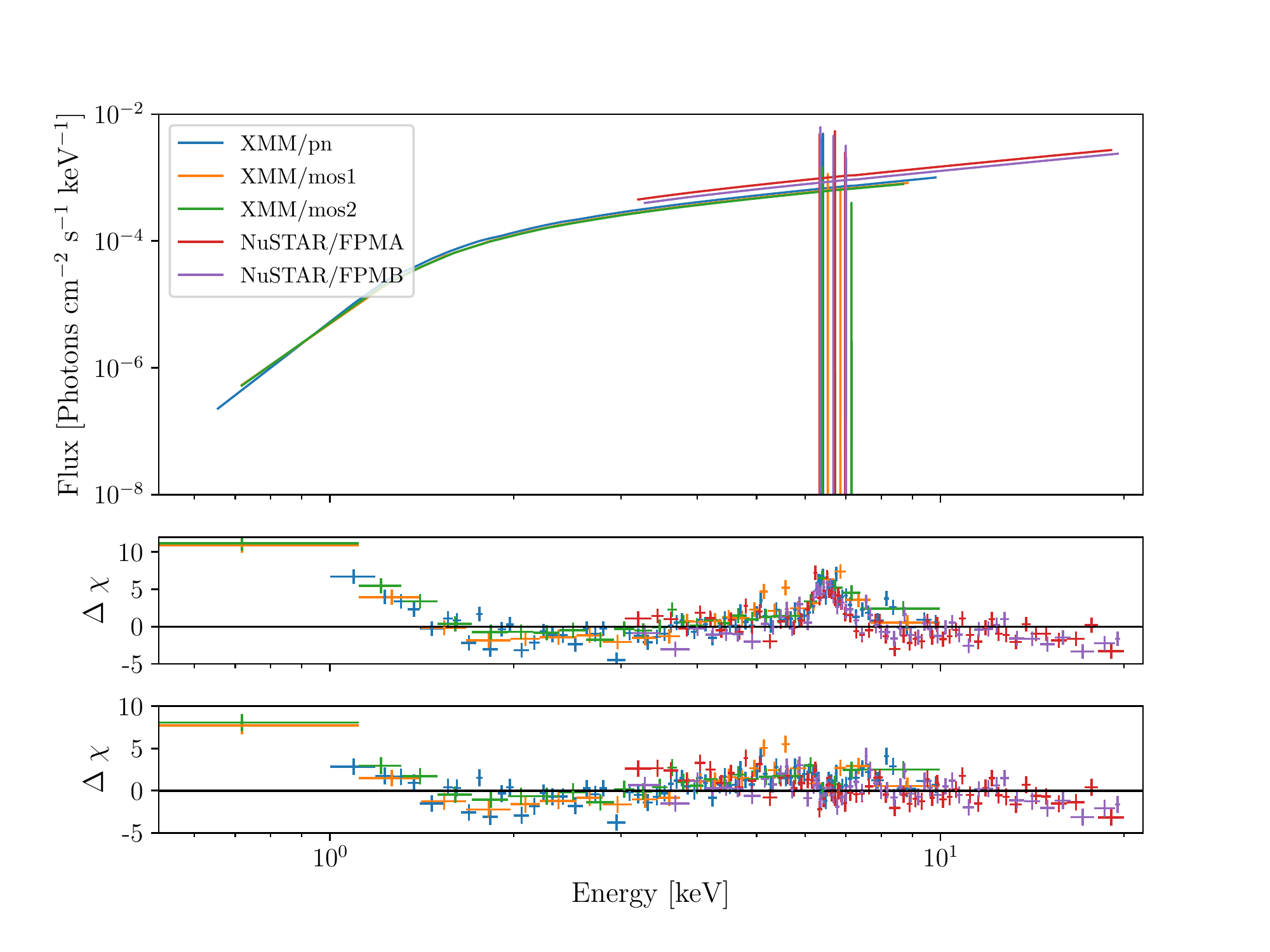}
    \caption{\xmm{} and \nustar{} spectral models are shown in the upper panels with the data residuals in the lower panel. (\textit{Top}) Model 2 in the top panel and residuals below it (\textit{Bottom}) Model 3 with associated residuals (top and bottom subpanels) as described in Tbl. \ref{tbl: spectral parameters} compared to \texttt{constant*tbabs(powerlaw)} residuals in the middle subpanel. Data are binned to a minimum of 5$\sigma{}$ significance for visual clarity.
    }
    \label{fig:xmm+nustar spectrum}
\end{figure}

We used the \nustar{} and \xmm{} data for our spectral fits, as these were the highest quality data sets. We first visually inspected the data, and found deviation from a smooth spectrum in the $6-7$~keV range. This is most significant in the PN spectrum (Fig. \ref{fig:pn Fe lines}).

The \xmm{} and \nustar{} spectra were first fitted separately with \textsc{xspec} with an absorbed powerlaw (PL) to identify general broadband spectral properties. A constant prefactor was used to allow for simultaneous fitting across all EPIC cameras (\xmm{}) or focal plane modules (\nustar{}). The prefactor was frozen to 1 in the first spectrum in the group, and all others were left free to vary. To check for spectral evolution, we first fit both datasets with a PL in their overlapping spectral range ($3-10$~keV); these fits yielded PL indices of $\Gamma{}=1.3_{-0.2}^{+0.3}$ in the \xmm{} data ($1\sigma{}$ uncertainties) and $\Gamma{}=2.2\pm{}0.2$ in the \nustar{} data ($1\sigma{}$ uncertainties). Therefore, \xabbr{} appeared to have varied significantly over the $\sim{}15$ day period between observations 5 and 6 in Tbl. \ref{tbl: observations}. However, large positive residuals remained in the  $6-7$~keV range, which encompasses the $F_{\rm e}$ line complex (right middle panel of Fig. \ref{fig:xmm+nustar spectrum}). We added a Gaussian component to account for the $F_{\rm e}$ lines. We refer to this absorbed PL plus Gaussian as Model 1. This model provided a good fit to both individual spectra with a reduced $\chi^{2}$=167.33/208 = 0.84 for \xmm{} and $\chi^{2}$=171.78/191=0.93 for the \nustar{} spectrum. All best-fit parameter values agreed within 1$\sigma{}$ in both instruments, therefore, we proceeded in fitting the two data sets together.

 We also used a prefactor for each spectrum in the joint fits with Model 1. The prefactors varied less than 10 percent for all \xmm{} instruments; we chose to freeze the PN factor to 1. For \nustar{}, the prefactors were best fit at 1.43 and 1.23 for Focal Plane Module (FPM) A and B, respectively; the prefactor for FPMA is $\geq{}$16\% greater than FPMB for all models tested. This discrepancy could be due to a rip in the multi-layer insulation recently reported by \cite{Madsen+2020}. 
 
 Motivated by the optical spectra that suggested that \xabbr\ was a CV (see below), and quite possibly an intermediate polar (IP), we introduced Model 2. Model 2 comprises two absorbing columns (tbabs for interstellar medium and pwab \citep{Done+Magdziarz1998} for a local partial covering absorber) with a broadband, isobaric cooling flow component (\texttt{mkcflow}) and a Gaussian for the $\sim{}$6.4~keV spectral line. While fitting Model 2, the nHmin parameter of pwab was frozen to 1.0$\times{}10^{15}$ cm$^{-2}$ and the lowT parameter was frozen to 0.0808 keV. The redshift parameter of \texttt{mkcflow} was frozen to the minimum 1.0$\times{}10^{-7}$, given the Gaia source distance of $\sim$2\,kpc. We show the best fit result in the left panel of Fig. \ref{fig:xmm+nustar spectrum}. Similar prefactor values were found for both Model 1 and 2. 
 
Finally, we fit the data with an absorbed PL with three Gaussians (referred to as Model 3), corresponding to the ionized $F_{\rm e}$ lines typically seen in CVs \citep{Hellier1998,Ezuka1999,Hellier2004}, and also suggested by the PN spectrum (Fig. \ref{fig:pn Fe lines}). These narrow line components were frozen at central energies 6.4, 6.7, and 6.97~keV and their standard deviations were frozen at 0.0~keV (one spectral bin width). This model is highlighted in the top right panel of Fig. \ref{fig:xmm+nustar spectrum}, with associated residuals shown in the bottom right. We compare all three models in Section \ref{sec: discussion}.

To obtain the source fluxes with each of these models, we froze the model normalizations and added a \texttt{cflux} component to the spectral model to measure unabsorbed flux. We set the energy band from 0.3~keV to 20.0~keV. The resulting values are reported in the last row for each model in Tbl. \ref{tbl: spectral parameters}. For spectral Model 2, we calculate a luminosity of $L_{0.3-20.0}=(2.44^{+2.43}_{-0.85})\times{}10^{33}$~erg s$^{-1}$, based on the unabsorbed flux and the Gaia counterpart distance. The flux and distance fractional uncertainties were added in quadrature; the distance uncertainty contributed to the large upper uncertainty of the luminosity.


\begin{table}[]
\begin{tabular}{llcc}
\hline
    Model         &       parameter        &  value  &  unit \\ 
                  &                        & $\pm{}$90\% confidence       &        \\\hline \hline
    {\bf 1. PL + Gaussian}         &     TBabs: ${N_{\rm H}}$           &  1.58$_{-0.25}^{+0.31}$&$10^{22}$ cm$^{-2}$  \\
  Cstat = 2308  &     PhoIndex &  1.02$_{-0.07}^{+0.07}$&           \\
    2812 d.o.f     &     PL norm     &  1.06$_{-0.14}^{+0.17}$& $10^{-4}$\\
or                &     LineE    &  6.50$_{-0.07}^{+0.05}$& keV\\
$\chi{}^{2}$= 2518.91  &     Gaussian Sigma    &  0.45$_{-0.09}^{+0.16}$& keV\\
   2812 d.o.f.     &     gaussian: norm     &  2.94$_{-0.41}^{+0.55}$&$10^{-5}$\\
$\chi{}^{2}_{red}$=0.90&unab. Flux$_{0.3-20.0}$ &  2.87$^{+0.16}_{-0.15}$&$10^{-12}$ erg s$^{-1}$ cm$^{-2}$ \\ \hline

{\bf 2. mkcflow + Gaussian}               &     Tbabs: ${N_{\rm H}}$          & 0.33$_{-0.16}^{+0.18}$ & $10^{22}$ cm$^{-2}$  \\
  Cstat = 2144    &     pwab: nHmax        & 29.05$_{-9.12}^{+21.07}$ &  $10^{22}$ cm$^{-2}$  \\
  2809 d.o.f.     &     pwab: beta         & -0.38$_{-0.07}^{+0.08}$ & \\
or                &     mkcflow: highT     &  63.18$_{-26.52}^{+16.72\rceil{}}$ & keV \\
$\chi{}^{2}$=2350 &     mkcflow: Abundanc  &  1.69$_{-0.90}^{+1.07}$ & \\
2809 d.o.f.       &   mkcflow: norm      &  8.85$_{-2.58}^{+8.39}$ & $10^{-12}$\\
$\chi{}^{2}_{red}$=0.84&     gaussian: LineE    &  6.41$_{-0.05}^{+0.08}$ & keV \\
                  &gaussian: Sigma    &  13.83$_{-8.03}^{+9.78}$ & $10^{-2}$ keV\\
                  &     gaussian: norm     &  1.23$_{-0.29}^{+0.44}$ & $10^{-5}$ \\
                  &unab. Flux$_{0.3-20.0}$ &  4.31$^{+0.47}_{-0.39}$&$10^{-12}$ erg s$^{-1}$ cm$^{-2}$ \\ \hline                  

       {\bf 3. PL + 3 Gaussians}         &     TBabs: ${N_{\rm H}}$          &  1.89$_{-0.27}^{+0.31}$& $10^{22}$ cm$^{-2}$  \\
  Cstat = 2335     &     powerlaw: PhoIndex &  1.08$_{-0.07}^{+0.07}$&           \\
    2812 d.o.f     &     powerlaw: norm     &  1.25$_{-0.15}^{+0.18}$& $10^{-4}$\\
                  &     gaussian: LineE    &  6.40                  & keV\\
or                &     gaussian: Sigma    &  0.0                   & keV\\
                  &     gaussian: norm     &  1.02$_{-0.14}^{+0.15}$&$10^{-5}$\\
$\chi{}^{2}$=2537  &     gaussian: LineE    &  6.70                  & keV\\
   2812 d.o.f.     &     gaussian: Sigma    &  0.0                   & keV\\
$\chi{}^{2}_{red}$=0.90&gaussian: norm     &  6.75$_{-1.40}^{+1.48}$&$10^{-6}$\\
                  &     gaussian: LineE    &  6.97                  & keV\\
                  &     gaussian: Sigma    &  0.0                   & keV\\
                  &     gaussian: norm     &  4.30$_{-1.20}^{+1.26}$&$10^{-6}$\\
                  &unab. Flux$_{0.3-20.0}$ &  2.94$^{+0.16}_{-0.16}$&$10^{-12}$ erg s$^{-1}$ cm$^{-2}$ \\
                  \hline
                  \hline
                  \vspace{-0.5cm}
\caption{Best fit parameters for Models 1, 2, and 3 (see NOTE)} 
\label{tbl: spectral parameters}
\end{tabular}

\tablecomments{Model 1: \texttt{constant*tbabs(powerlaw+gaussian)} \\
Model 2: \texttt{constant*phabs*pwab(mkcflow + gaussian)} \\ 
Model 3: \texttt{constant*Tbabs(powerlaw + gaussian + gaussian + gaussian)}. \\
{Brackets on parameters denote that the uncertainty extends to the upper ($\rceil{}$) or lower limit ($\rfloor{}$) of the model parameter allowed by \textsc{xspec}.}}

\end{table}

\subsubsection{Optical spectra} 




\begin{figure}[!t]
    \centering
    \includegraphics[angle=270, width=0.9\textwidth]{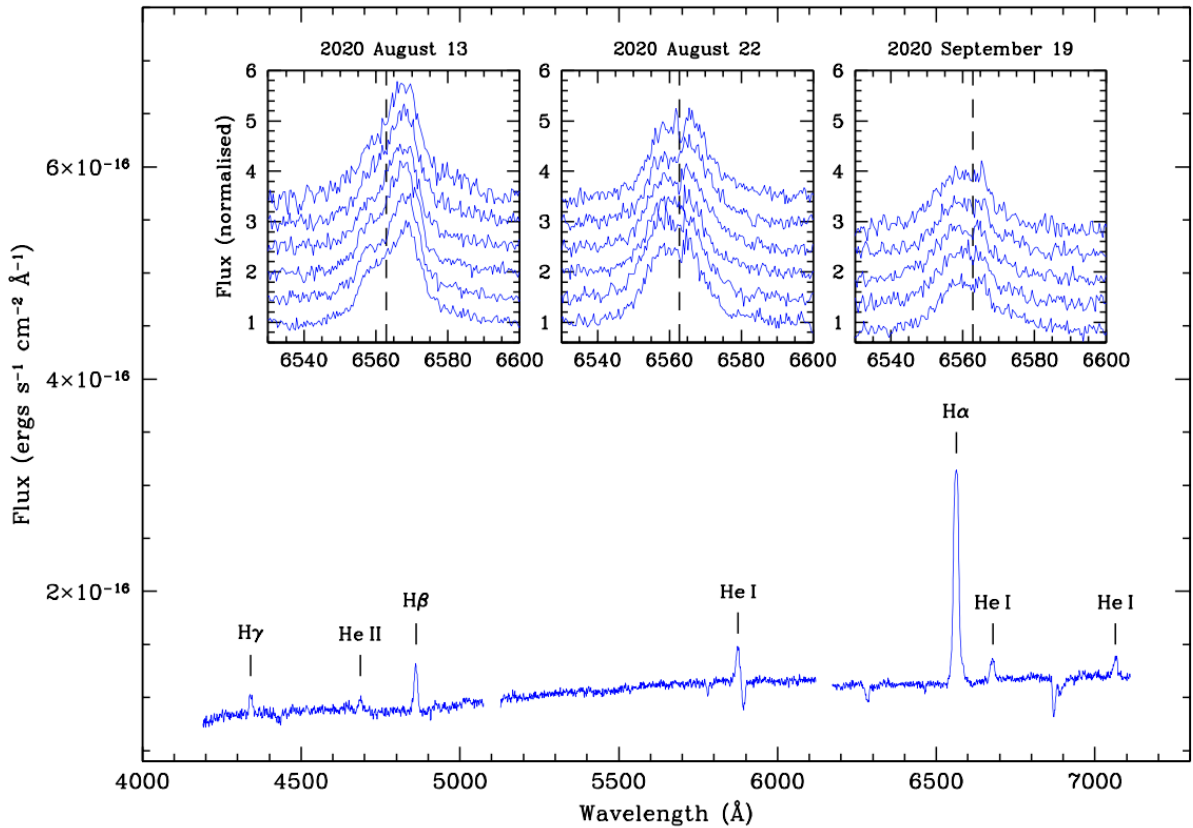}
    \caption{The average SALT RSS spectrum of J1839 covering the wavelength range of 4190 to 7100 {\AA}. The main emission lines are marked and labelled. The insets show the varying H$\alpha{}$ emission line profiles as observed at higher spectral resolution (PG1800) on SALT on three different days; each sequence shows a series of consecutive 500-s spectra (first to last: bottom to top). The vertical dashed line in the insets indicates the rest wavelength of H$\alpha{}$.}
    \label{fig:IR spectroscopy}
\end{figure}


Fig.~\ref{fig:IR spectroscopy} shows the averaged flux-calibrated SALT RSS spectrum of J1839, where we combined the two highest signal-to-noise spectra obtained on 2020, July 21 and 23, respectively. The spectrum is characterised by strong H I (Balmer), He I and He II emission lines, on top of a continuum which rises to the red. This spectrum establishes \xabbr\ as a likely CV. The \textsc{H}$\alpha{}$ emission is the strongest and we also see both the high excitation He II 4686\AA{} and Bowen fluoresence ($4640-4660$\AA) lines, both a common feature seen in magnetic CVs as a result of the ionizing soft X-ray emission \citep{McClintock1975,Mukai2017}. Whereas the He II 4686\AA{} line can be brighter than the neighboring H$\beta$ line in polars, it is typically weaker in intermediate polars.  

Given the strength of the H$\alpha{}$ line, we obtained higher resolution SALT spectra (using the PG1800 grating on RSS) on three different nights in 2020 August - September (Obs. 27, 30 \& 31 in Tbl.~\ref{tbl: observations}). The inserts in Fig.~\ref{fig:IR spectroscopy} show the varying H$\alpha{}$ profiles. From night to night there is clear variability in the line profile but unfortunately the coverage of the high resolution spectra is too sparse to look for coherent variability on the time scale of a possible orbital period. 

\subsection{Timing}



\begin{figure}[!t]
    \centering
    \includegraphics[width=0.9\textwidth,draft=false]{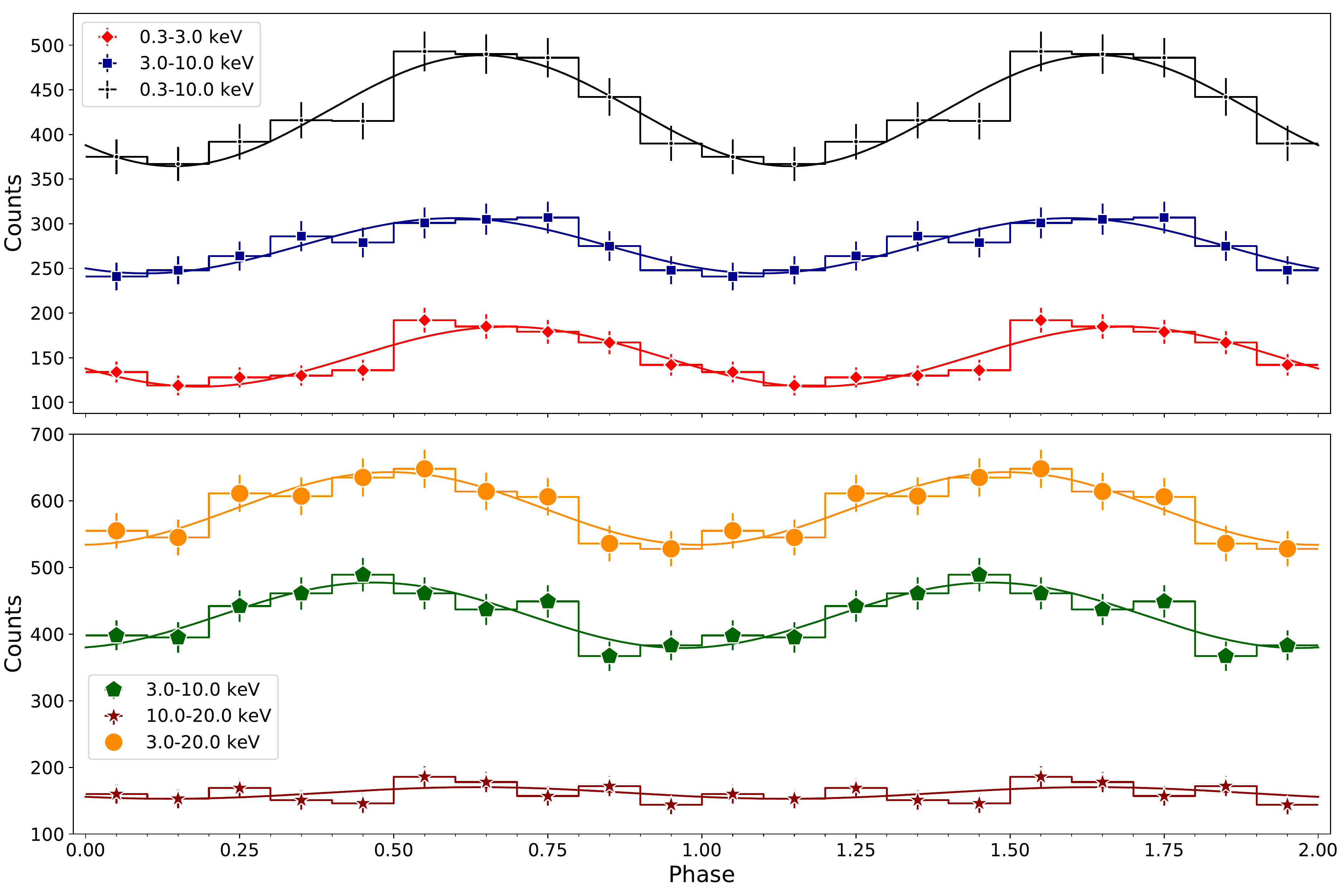}
    \caption{Pulse profile using (\textit{Top}:) the combined \xmm{} EPIC cameras in three energy bands from Obs. 5 and (\textit{Bottom}:) the combined focal plane modules of \nustar{} in three energy bands. The data plotted are repeated over two full periods for visual clarity, in bins of 0.1 period $\sim{}$44.97~s.}
    \label{fig:X-ray Period}
\end{figure}
\begin{figure}[!b]
    \centering
    \includegraphics[angle=270, width=1\textwidth]{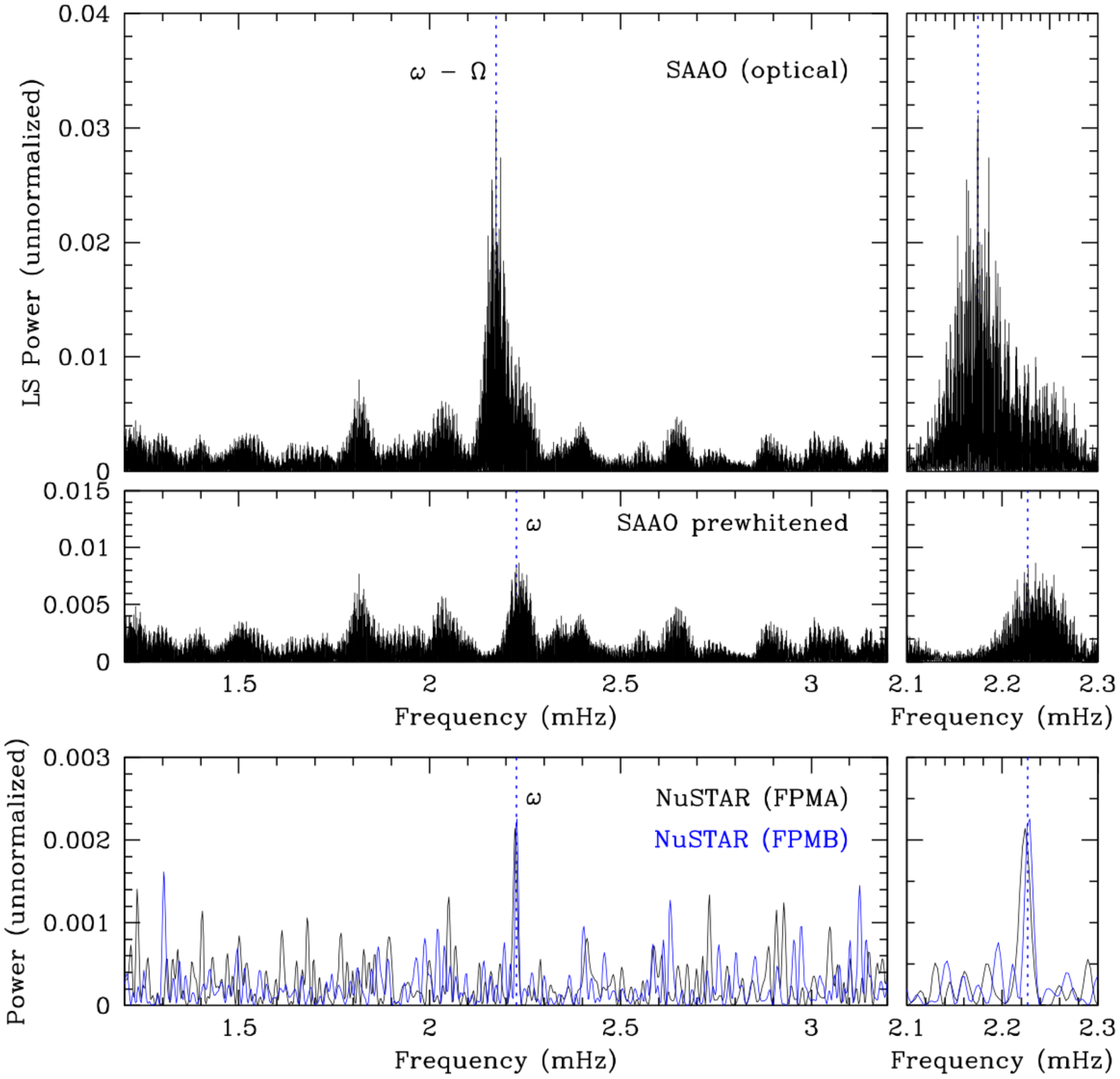}
    \caption{Lomb-Scargle periodograms of the optical data from SAAO (top and middle panels) and the X-ray data from \nustar{} (lower panels). The left panels show the periodogram over a wide frequency range (1.2 - 3.2 mHz), whereas the right panels zoom in on the peaks associated with the spin and the beat frequencies over a narrow frequency range (2.1 - 2.3 mHz). The middle panels show the residual signal after prewhitening with the dominant beat frequency ($\omega - \Omega$) identified in the top panel.}
    \label{fig:LSsaao}
\end{figure}

We searched the \xmm, \nustar, and \saao{} data for periodicities. 
For \xmm{} we created a Z$^{2}_{1}$ statistic \citep{Buccheri+1983}, in the $0.3-10.0$\,keV band of EPIC PN. The resulting power spectrum showed a maximum at $(2.228\pm0.004)\times{}10^{-3}$~Hz, i.e., a period of $448.7\pm0.3$\,s. 
To search for periodicities in the \nustar{} data, we corrected the photon arrival times to the Solar System barycenter using the \textsc{heasoft v.6.23} command \texttt{hdaxbary}. We set the location to the source centroid measured with \chandra{}. We extracted source events from the 80\% Enclosed Energy Fraction (EEF) radius of 75\arcsec{} for the most conservative $3-4.5$\,keV energy range \citep[][Fig. 5]{An+2014}.
A Lomb-Scargle frequency analysis of the \nustar{} data (FPM A and B) shows a peak at $f_{NuSTAR1}=(2.2268\pm0.0016) \times{}10^{-3}$~Hz, consistent with the peak frequencies observed in the XMM observations, but with considerably better accuracy. The resulting power spectrum is shown in the lower panel of Fig.~\ref{fig:LSsaao}. The X-ray period is $449.1\pm0.3$\,s. A Z$^{2}_{1}$ test in the combined FPMA and FPMB \nustar{} data within a $3-30$\,keV range, found a $5\sigma$ peak at $f_{NuSTAR2}=(2.2239\pm0.0017) \times{}10^{-3}$~Hz, corresponding to an X-ray period of $449.7\pm0.6$\,s.
Given that the above period values are all consistent within $\sim1\sigma$, and also taking into account the extended energy range of the latter estimate, as well as the significant jitter associated with white dwarf timing, we adopt the latter as the spin period of \xabbr{}, as it is far too short to be the orbital period of a non-degenerate hydrogen-rich CV. \citet{Littlefield2016} demonstrated a pronounced (almost 0.2 cycle) phase shift in the optical spin pulse of FO Aqr when it entered a low state (see their Figure 5). \xabbr{} may also be demonstrating phase jitters of up to 0.1 cycle in total range during the source high state in figures \ref{fig:X-ray Period} and \ref{fig:LSsaao}, which can be plausibly interpreted as due to accretion rate changes. While phase jitters have not been directly observed in X-rays, \citet{Hellier1997} observed orbital phase shift in the eclipse egress timing in XY Ari, which they interpret as random shifts in the position of the accreting spots in that IP.

We then folded all three EPIC camera lightcurves with the 449.7\,s period, (Fig. \ref{fig:X-ray Period}) using the \texttt{fold\_events} command in Python package \textsc{Stingray v.0.1} \citep{Huppenkothen+2019}. The folded pulse profiles were fit with the first harmonic function of \citet{Bildsten+1997}, i.e., \texttt{A}sin(2$\pi{}\phi{}$)$+$\texttt{B}cos(2$\pi{}\phi{}$)$+$\texttt{C}, with parameters \texttt{A}, \texttt{B}, and \texttt{C} free to vary. The resulting best-fit parameters, their standard deviations ($\sigma{}_{X}$), and $\chi{}^{2}$ fit statistic are reported in Tbl. \ref{tbl: timing parameters}. 

We calculated the root mean square pulsed fraction ($PF_{RMS}$) using Eq. \ref{Eq. pulsed fraction}. 

\begin{equation}\label{Eq. pulsed fraction}
    PF_{RMS}=\texttt{C}^{-1}\sqrt{0.5\times{}\left(\texttt{A}^{2}+\texttt{B}^{2}-\sigma{}_{\texttt{A}}^{2}-\sigma{}_{\texttt{B}}^{2}\right)}
\end{equation}

To calculate the uncertainty on $PF_{RMS}$, we performed $10^4$ simulations of the pulse profile. In each simulation we created a synthetic pulse profile by drawing data points from Gaussian distributions. The Gaussian was unique to each phase bin, with a mean equal to the bin value and standard deviation equal to the bin uncertainty. Each synthetic pulse profile was fit with the same model described above, and the pulsed fraction was recorded. After all simulations were complete, we created a histogram of $PF_{RMS}$ values, which was fit with a Gaussian. The standard deviation of the best-fit Gaussian was taken to be the uncertainty of $PF_{RMS}$. 
We used the same methods as in the \xmm{} data to plot and determine the pulsed fraction in the \nustar{} data.

\begin{table}[]
    \centering
    \begin{tabular}{|l|c|c|c|c|c|c|}
    \hline 
	Parameter [units]&\multicolumn{3}{c|}{Combined XMM EPICs} &\multicolumn{3}{c|}{Combined NuSTAR FPMs} \\ 
	\hline 
	\hline 
							&  0.3--3.0 keV&  3.0--10.0 keV & 0.3--10.0 keV  & 3.0--10.0 keV & 10.0--20.0 keV  & 3.0--20 keV\\ 
	\hline 
	
	Constant, \texttt{C} [cts]& 151.20 $\pm{}$ 3.45 & 275.40 $\pm{}$ 3.11 & 426.60 $\pm{}$ 5.22 & 428.20 $\pm{}$ 5.73  & 161.60 $\pm{}$ 4.49 & 588.50 $\pm{}$ 5.43  \\ 

	\hline 
	
	Amplitude \texttt{A} [cts]& -30.82 $\pm{}$ 4.87 & -17.84 $\pm{}$ 4.40 & -48.66 $\pm{}$ 7.38 & 9.67 $\pm{}$ 8.11  & -6.53 $\pm{}$ 6.35 & 2.19 $\pm{}$ 7.68 \\
	\hline
	
	Amplitude \texttt{B} [cts]& -13.30 $\pm{}$ 4.87 & -25.30 $\pm{}$ 4.40 & -38.60 $\pm{}$ 7.38 & -48.13 $\pm{}$ 8.11 & -5.80 $\pm{}$ 6.35 & -54.50 $\pm{}$ 7.68 \\ 
	\hline
	
	Pulsed Fraction [\%] & 15.0 $\pm{}$ 3.0 & 7.6 $\pm{}$ 2.3 & 10.0 $\pm{}$ 1.7 & 7.9 $\pm{}$ 2.4 & $\dagger{}$ & 6.4 $\pm{}$ 1.9 \\ 
	\hline
	
	$\chi{}^{2}$ ($\chi{}^{2}_{RED}$; 7 DOF) & 5.37 (0.77) & 2.42 (0.35) & 4.53 (0.65) & 4.24 (0.61) & 6.63 (0.95) & 2.76 (0.39) \\ 

	\hline
    \end{tabular} 
    \tablecomments{$\dagger{}$ The pulsed fraction could not be determined, due to a low number of pulsed counts.}
    \caption{Best fit parameters for the pulse profile model \texttt{C}$+$\texttt{A}\texttt{sin(}$2\pi{}\phi{}$\texttt{)}+\texttt{B}\texttt{cos(}$2\pi{}\phi{}$\texttt{)} are shown for energy bands from observations 5 and 6. Errors are at the 68\% confidence interval.}
    \label{tbl: timing parameters}
\end{table}


 The 2020 SAAO high speed photometry (see Sect. 3.4.1) was used to search for coherent short-period optical modulations. We employed the generalized Lomb-Scargle period search as implemented in {\tt{VARTOOLS}} \citep{Hartman2016} to identify peak frequencies in the optical data. The top panel of Fig.~\ref{fig:LSsaao} shows the Lomb-Scargle periodogram of the SAAO photometry between 1.2 and 3.2 mHz (left side), and a zoomed in view between 2.1 and 2.3 mHz (right side). A clear peak is identified at $2.17441 \times 10^{-3}$ Hz with a formal false alarm probability of $1.1 \times 10^{-14}$, which we associate with the beat frequency ($\omega - \Omega$), where $\omega$ is the spin frequency determined from the X-ray data, and $\Omega$ is the inferred orbital frequency. The optical beat frequency corresponds to a period of 459.89 $\pm$0.13\,s. When interpreted using the $449.7\pm 0.3$ X-ray spin period, this implies that the inferred orbital period is $5.6 \pm 0.6$ hr, 
where the error is dominated by the uncertainty in the X-ray spin frequency. 

After pre-whitening the optical data with the beat frequency, a small residual remains at the location of the X-ray spin frequency. This is shown in the middle panels of Fig.~\ref{fig:LSsaao}. The zoomed in view (middle-right panel) shows a number of alias peaks around the X-ray spin frequency (vertical dashed line), making it hard to improve on the exact determination of the spin frequency. Nonetheless, there is clearly some power in the optical Lomb-Scargle periodogram at the spin frequency. The simultaneous presence of spin and beat periods in the optical is typical of IPs \citep{Warner1986,Pretorius2009,Littlefield2016}. The relative strength of the beat modulation indicates that optical pulses are dominated by the X-rays reprocessed by structures fixed in the binary frame, such as the secondary or the bright spot where the accretion stream interacts with the accretion disk.

 In general, the SAAO photometric runs are less than 5 hr in length, with only one long observing run of 7 hrs (Obs. 22). This, together with the complex alias structure in the power spectrum, dominated by the $\sim$5 d sampling, makes it hard to reliably identify low frequency orbital modulations in the SAAO data. However, two period maxima are seen in the power spectrum, close to the purported $\sim 5.6 \pm 0.3$ h orbital modulation, at 5.25 and 5.49 d. We searched for longer periods using the more extensive coverage ZTF photometry of \xabbr{} in the r-band
and found a peak in the power spectrum at 5.448 h, consistent with the predicted orbital period, although similar power is seen at several of the 1 cycle d$^{-1}$ aliases either side. 



\subsection{Photometry} \label{sec:photometry}

We searched with the VOSA tool \citep{Bayo+2008} for archival data within 2\arcsec{} of the \chandra{} position of \xabbr{}. We found a candidate counterpart in the Panoramic Survey Telescope and Rapid Response System \citep[Pan-STARRS,][]{Chambers+2016} g, r, i, z, and y bands, PSO J183919.988$-$045353.099 (PS1 identifier 102122798332882651), which is 0.178\arcsec{} offset from \xabbr{}. We then searched the Gaia data release 2 \citep{Gaia+2018} and found G, G$_{BP}$, and G$_{RP}$ magnitudes of the source Gaia DR2 4256603449854150016, which is offset by 0.193\arcsec{}  from the \cxo{} position of \xabbr (see also Fig. \ref{fig: cxoposition}). 

\subsubsection{SAAO high speed photometry}



\begin{figure}[!b]
     \centering
     \includegraphics[trim={0 3cm 0 3cm}, angle=270, width=0.8\textwidth]{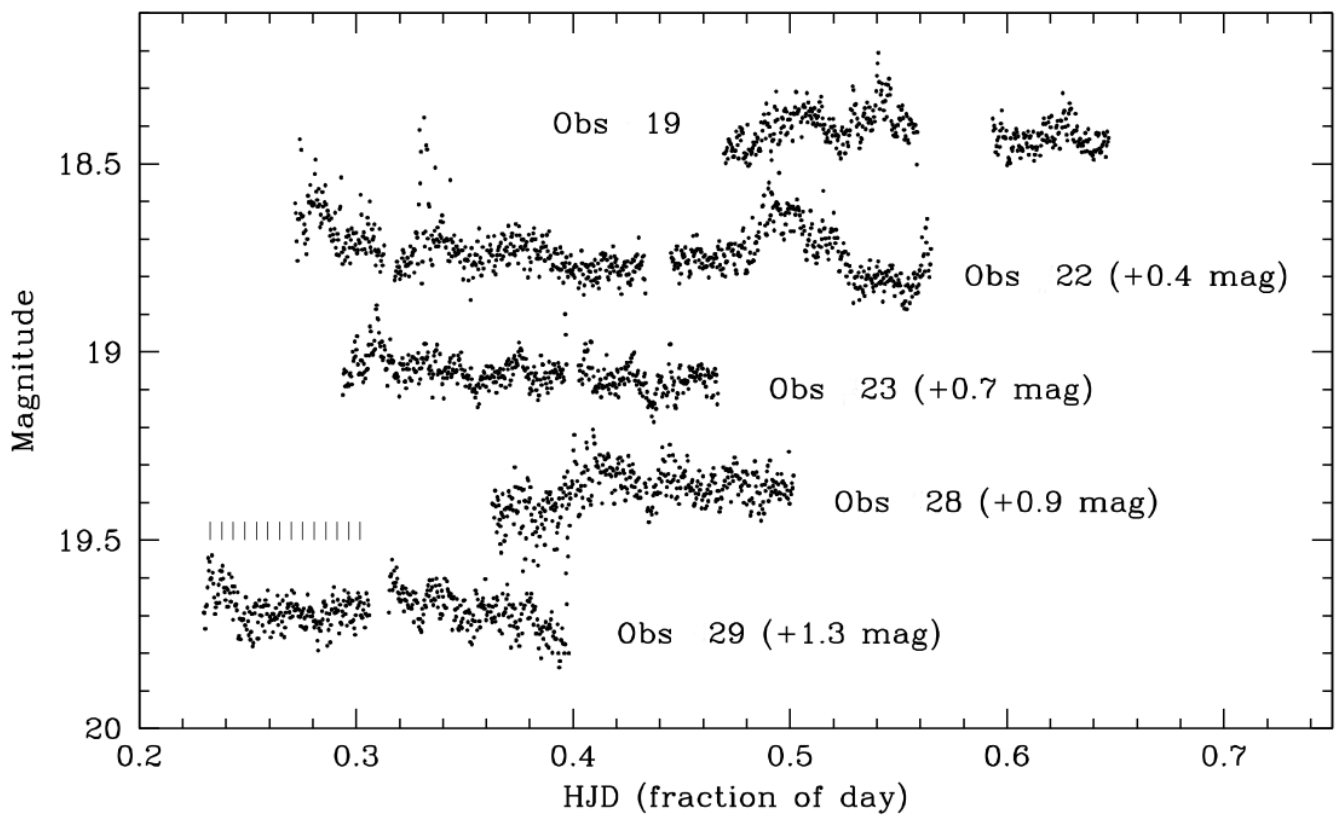}\
     \caption{Light curves of \xabbr{} taken with SAAO 1-m telescope. Obs. 19 is shown at the correct brightness, whereas subsequent observing runs have been displaced vertically for display purposes only. The data are calibrated to the Pan-STARRS r-band. The vertical bars for Obs. 29 indicate the expected positions of the spin modulated optical pulses.}
     \label{fig:LCsaao}
 \end{figure}

Fig.~\ref{fig:LCsaao} shows the SAAO high speed photometry of J1839 for five of the longest observing runs in 2020. The data were obtained in white light (unfiltered) and calibrated using Pan-STARRS r-band photometry of reference stars in the field. Such a calibration is only accurate to $\sim$ 0.1 mag, see discussion in \citet{Coppejans+2014}. J1839 is around 18.4 mag during these observations. In Fig.~\ref{fig:LCsaao}, Obs. 19 is shown at the correct brightness and subsequent runs have been offset for display purposes. The light curves show the characteristic variability of a CV. In parts of the light curve, the optical modulation at 459.89 s is clearly visible by eye, e.g in Obs. 29 where we have marked the peaks of the beat period by small vertical dashes in Fig.~\ref{fig:LCsaao}. No clear orbital modulation on time scales of $\sim$ 5 hr is visible in the light curves.

 \subsubsection{Optical polarization} \label{sec:polarization}
\begin{figure}[!b]
     \centering
     \includegraphics[trim={0 6cm 0 6cm}, width=0.99\textwidth]{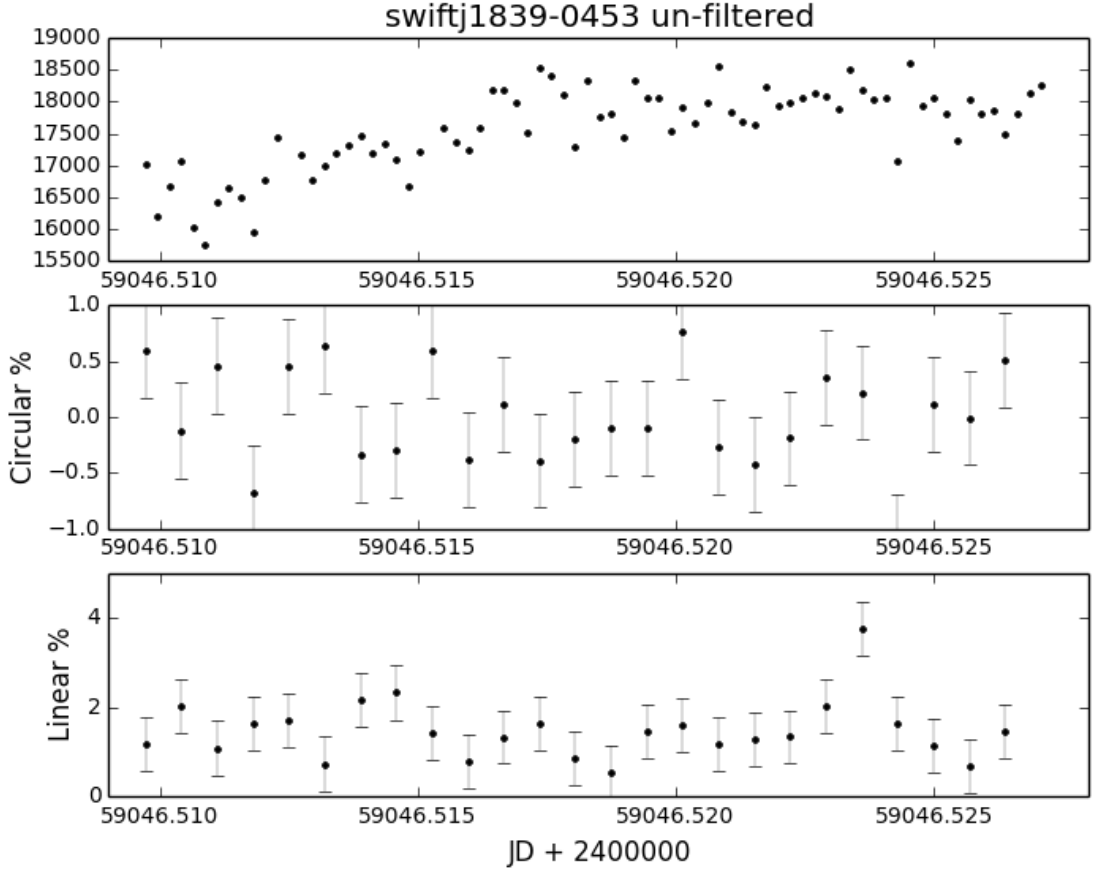}\
     \caption{HIPPO photopolarimetry of \xabbr{} taken with SAAO 1.9-m telescope (Obs. 21). The top panel shows total counts and the next two panels show the linear and circular polarization, respectively.}
     \label{fig:pol}
 \end{figure}

The results of the HIPPO photopolarimetry of \xabbr{} are presented in Fig.~\ref{fig:pol}. No circular polarization, typical of magnetic CVs, particularly polars, was detected ($<\it{V/I}>$ = 0.01~\% S.D. = 0.46~\%). Since most IPs do not show detectable polarization, this is not surprising. The mean linear polarization value was $<p>$ = 1.47~\% with S.D. = 0.67\%, with no obvious variability. This is consistent with an interstellar origin. The position angle of the linear polarization is $\theta$ = 95${^\circ} \pm 12^{\circ}$, consistent with an ISM origin.

\subsubsection{UV and Optical}
For the \swift{}/UVOT we used the command \texttt{uvotsource} in the \texttt{heasoft v6.23} to determine the source AB magnitude (uncorrected for extinction); these are reported in column 6 of table \ref{tbl: uvot_results}.

In the \xmm{}/OM, we used the SAS routine \texttt{omdetect} (method 1) to find counterparts in the U and UVW1 filters, which were detected at 26.89$\sigma{}$ and 8.11$\sigma{}$ significance, respectively. These candidate counterparts are 1.10\arcsec{} and 1.29\arcsec{} from the \chandra{} location, respectively. Both of the sources have nonzero quality and confusion flags, which we address below. The latter is expected in crowded regions like the Galactic plane. 

The UVW1-band source is within an area of enhanced emission (possibly diffuse) and has a bad pixel (QFLAG=33). There were also one or more sources detected at a distance of 6$-$12 pixels (CFLAG=1). Source confusion was checked visually, and the \texttt{omdetect} region centroid is centered on the source without significant light from neighboring sources.

The U-band source has the same flags as the UVW1-band, however, it also lies near a bright source (total QFLAG=97) and has one or more sources 3-6 pixels away (total CFLAG=3). The former flag is quite possibly due to the increased sensitivity and spectral breadth compared to the UVW1-band as well as the source lying within the galactic plane.\footnote{See \href{https://xmm-tools.cosmos.esa.int/external/xmm_user_support/documentation/uhb/omfilters.html}{https://xmm-tools.cosmos.esa.int/external/xmm\_user\_support/documentation/uhb/omfilters.html}} Similarly, the source region was visually inspected and deemed acceptable with little neighbor contamination.

\begin{table}[!t]
\begin{tabular}{llcccc}
\hline
Obs. & ID & Telescope    & Start Time [UT]      & Filter & Magnitude (or 5$\sigma{}$ limit) \\
     &    &              & [dd Mmm yyyy hh:mm]  &        & [AB mag]  \\ \hline\hline

1.  & 00044416001 & UVOT & 21 Mar. 2013 06:59   & UVM2 & $>$21.22\\
2.  & 00087393001 & UVOT & 13 Jul. 2017 06:02   & UVW1 & $\dagger{}$ \\
3.  & 00087393002 & UVOT & 14 Nov. 2017 02:42   & UVW1 & $\dagger{}$\\
4.  & 00010900001 & UVOT & 01 Oct. 2018 17:09   & U    & 19.70$\pm{}$0.15$\pm{}$0.02 \\
5.  & 0821860201  & OM & 18 Oct. 2018 11:27     & UVW1 & 19.67$\pm{}$0.05\\
    &             &    &                        & U    & 18.43$\pm{}$0.02\\
7.  & 00088814001 & UVOT & 02 Nov. 2018 12:30   & U & 20.28$\pm{}$0.10$\pm{}$0.02\\
8.  & 00088814002 & UVOT & 08 Nov. 2018 13:34   & UVM2 & $>$21.11\\
10. & 00087393003 & UVOT & 07 Jul. 2019 21:29   & UVW1 & $\dagger{}$\\

\hline\hline
\end{tabular}
\caption{Observations of \xsrc{} by UVOT.}
\tablecomments{\swift{} magnitude uncertainties shown are statistical first (when available) then systematic uncertainty. $\dagger{}$:~Source projected out of the field of view}
\label{tbl: uvot_results}
\end{table}

\begin{figure}[!b]
    \centering
    \includegraphics[scale=0.7, draft=false]{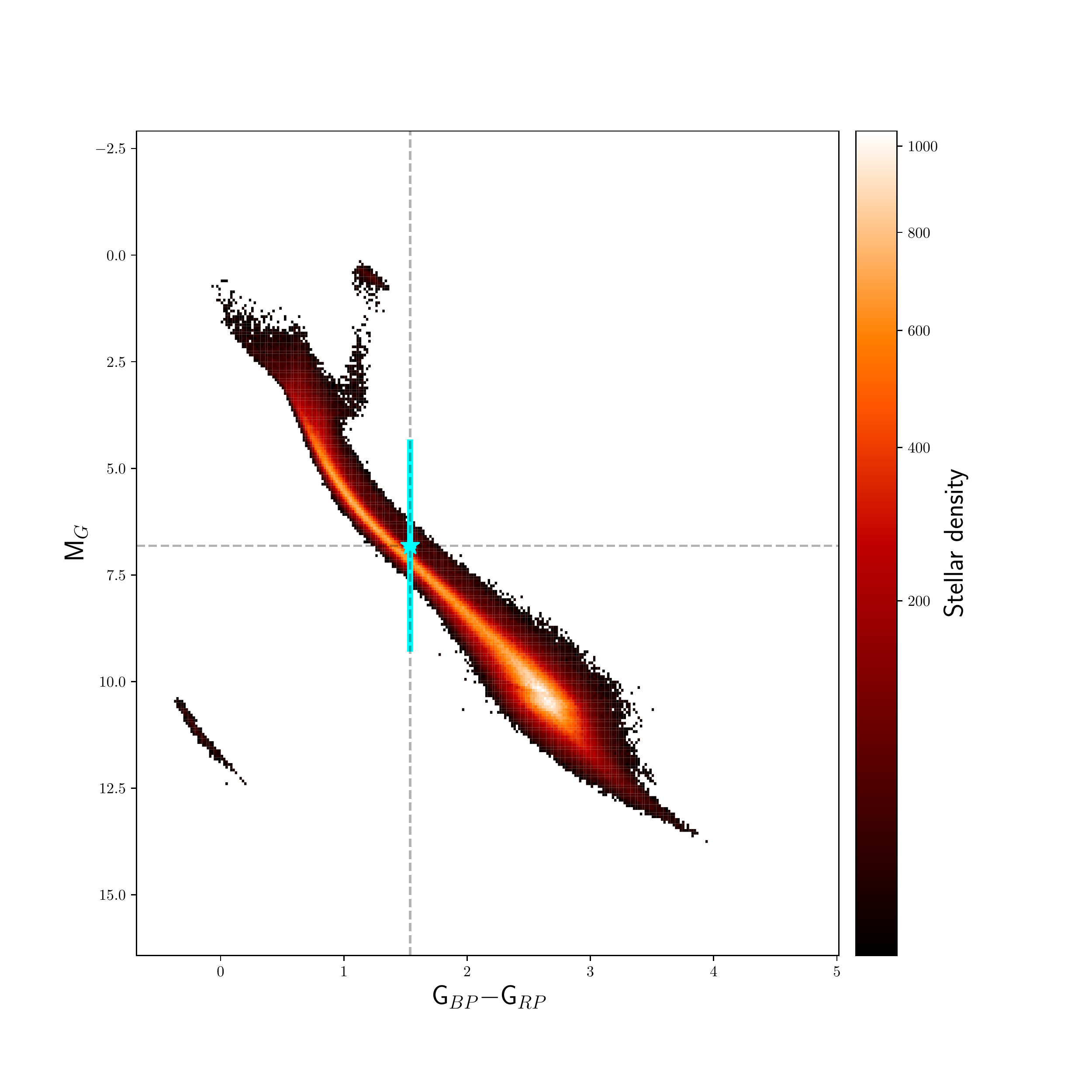}
    \caption{The location of \xabbr{} (cyan) on the Gaia H-R Diagram, using sources $\leq{}$200~pc from the sun. Other filtering criteria as in \citet{Ruiz-Dern+2018}.}
    \label{fig:Gaia HR Diagram}
\end{figure}

    \label{fig:radio images}

\subsubsection{Radio}
We find no radio sources at or near the \chandra{} position. The closest source, an unidentified radio triple, is $\sim{}30$\arcsec{} away. The 3$\sigma{}$ upper limits (i.e., 3$\times{}$rms noise) at the \chandra{} position are 0.043 mJy at 9.746~GHz for Obs. 15; 0.038 mJy at 9.746~GHz for Obs. 16; and 0.028 mJy at 9.370~GHz for Obs. 17. We combined the data from Obs. $15-17$ and obtained a 3$\sigma{}$ upper limit of 21.6~$\mu{}$Jy at an effective central frequency of 9.37~GHz with a 2.75~GHz bandwidth.




\begin{table}
        \begin{tabular}{|l|c|c|}
                \hline
                Parameter          &       Value                          &  Citation\\\hline \hline
                Luminosity$_{0.4-20\;keV}$         & $2.71^{+2.69}_{-0.91} \times{}10^{33}$~erg s$^{-1}$ & This Work \\\hline
                $P_{s}$& $449.7$~s & This Work \\\hline{}
                $P_{o}$& 5.6 $\pm{}$ 0.3~hr   & This Work \\\hline{}
                Distance           & $2.3^{+1.9}_{-0.8}$~kpc &  \cite{Bailer-Jones+2018}\\\hline{}
                $\mu{}_{\alpha{}}$ & -1.1 $\pm{}$ 0.4~mas yr$^{-1}$   &  \cite{GaiaCollab2018}  \\\hline{}
                $\mu{}_{\delta{}}$ & -2.6 $\pm{}$ 0.4~mas yr$^{-1}$   &  \cite{GaiaCollab2018}  \\
                \hline
        \end{tabular}
        \caption{Fundamental properties of \xabbr{}.}
    \label{tbl: summary parameters}
\end{table} 


\section{Discussion} \label{sec: discussion}



The intrinsic properties of \xabbr{} derived above and summarized in Table \ref{tbl: summary parameters} are typical of known IPs: the X-ray luminosity is typical \citep{Pretorius+2014}, and the measured spin period and the inferred orbital period are both well within the normal range for IPs \citep{Mukai2017}. Further, the success of our X-ray spectral modeling shows that the X-ray emission may originate in hot plasma, possibly having a multi-temperature nature, that is located next to a cold surface containing Fe of low ionization states. As to the optical magnitude and color, we must first account for interstellar reddening. The 3-d extinction map of \cite{Lallement+2019}\footnote{Available online at https://stilism.obspm.fr/} only extends to 2.25 kpc in this direction, and gives an extinction of E$_{B-V}$=1.08$\pm$0.45 mag at that distance. The lower end of this range may be appropriate for \xabbr, since  the {\sc mkcflow} fits to the X-ray spectrum suggests an interstellar N$_H$ of 3.3$^{+1.8}_{-1.6} \times 10^{21}$ cm$^{-2}$, or E$_{B-V}\approx$ 0.5. Given these uncertainties, it is impossible to be precise, but the location of \xabbr\ in Fig. \ref{fig:Gaia HR Diagram} will shift up and to the left, likely moving it into a location occupied by the majority of IPs as compiled by \cite{Abril+2020}. Thus, \xabbr\ appears to be a typical IP seen at a larger than typical distance, thus extending the reach of the CV population in our Galaxy.




The discovery of J1839 serves as a reminder that deeper X-ray surveys of the Galactic plane will likely yield detection of an increasing number of IPs. The large majority of currently known IPs have been discovered through X-ray surveys using collimated (non-imaging) instruments, such as Uhuru and HEAO-1, and using coded-mask aperture instruments, Swift BAT and INTEGRAL, perhaps out to distances of 1--2 kpc \citep{Mukai2017}. 


The work presented here also demonstrates the feasibility of detecting IPs far away from the Solar neighborhood in a systematic manner. While a number of even more distant IPs have been reported, those close to Galactic center distances are not suitable for detailed multi-wavelength studies such as the one we have performed. Moreover, very distant IPs discovered serendipitously can not be reliably used for statistical population studies. Future DGPS discoveries of IPs have the potential to lead to a direct observational study of the Galactic distribution of CVs, a subject rarely discussed in literature, with a few exceptions, e.g., \cite{Britt+2015}. Such studies will place the estimates of the CV contributions to the Galactic Ridge on a much firmer ground \citep{Mukai+1993}.

Deep X-ray surveys might also uncover a relatively nearby population of low luminosity IPs (LLIPs), if such objects exist \citep{Pretorius+2014}. Current indications are that there is a deficit of IPs with X-ray luminosities in the 10$^{32}$ to 10$^{33}$ ergs\,s$^{-1}$ range, but perhaps with a separate population of LLIPs below 10$^{32}$ ergs\,s$^{-1}$. DGPS and other future X-ray surveys have the potential to confirm or refute this possibility.

\section{Conclusions} \label{sec: conclusions}
We described the multi-wavelength followup of the source \xabbr{}, discovered within the scope of the DGPS. We found source counterparts in Gaia and PanSTARRS data which allowed us to determine source location, distance, and proper motion. We detect counterparts to \xabbr{} in the UV, optical, and IR regimes, however no counterpart was found in the radio S or X bands in VLA archival data and our followup observations, respectively.

We found a hard X-ray spectrum with three $Fe$ lines. This is fit well with a \texttt{mkcflow} spectral model with an additional Gaussian line for neutral $F_{\rm e}$, which is doubly-absorbed by local and ISM material. We find H$_{\alpha{}}$ and He lines in the optical/IR spectra, 
the former of which are double-peaked and evolves with time. 

In our timing analysis we found two periods, at 449.7~s and 459.89~s which we interpret to be the spin and beat periods of an intermediate polar (IP) magnetic CV system. From these periods we infer an orbital period of 5.6~h 
This orbital period is consistent with periods seen with moderate significance in two independent datasets of long-baseline optical photometry (ZTF and  SAAO). The orbital, spin and beat period detections provide strong evidence of an intermediate polar (IP) classification for \xabbr{}. Other supporting evidence includes the complex nature of the optical emission lines and the presence of He II 4686\AA~ emission. Furthermore, we find a sinusoidal pulse profile in the X-ray data at the spin period, with a pulse fraction that decreases with increasing energy, another characteristic that supports the IP origin of J1839.

\facilities{\swift{}, \nustar{}, \xmm{}, \chandra{}, SALT, SAAO, Gaia, Pan-STARRS, ZTF, APO, VLA}
\software{SAS, Pydis, Pyspeckit, Stingray, DS9, Python 3, Heasoft, CASA}

\bibliography{main.bib} 

\vspace{0.4cm}
Acknowledgements:
The authors wish to thank Sylvia Rose Kowalski and Deena Mickelson for their roles in VLA data acquisition. They also thank Hannes Breytenbach for taking some of the SAAO data. PAW acknowledges financial support from the University of Cape Town and the National Research Foundation. The SALT observations reported here were obtained through the SALT Large Science program 2018-2-LSP-001, with D.B. as PI, who also acknowledges support of the National Research Foundation. C.K., and N.G. acknowledge support under NASA Grant 80NSSC19K0916 and Smithsonian Astrophysical Observatory Grant GO9-20057X. B.O. is supported in part by the National Aeronautics and Space Administration through grants NNX16AB66G, NNX17AB18G, and 80NSSC20K0389.

We thank the NRAO for the generous allocation of VLA time for our observations. The National Radio Astronomy Observatory is a facility of the National Science Foundation operated under cooperative agreement by Associated Universities, Inc.

This work has made use of data from the European Space Agency mission Gaia (https://www.cosmos.esa.int/gaia), processed by the Gaia Data Processing and Analysis Consortium (DPAC, https://www.cosmos.esa.int/web/gaia/dpac
/consortium). Funding for the DPAC hasbeen provided by national institutions, in particular the institutions participating in the Gaia Multilateral Agreement.

\end{document}